# Laser Shock Peening in Hydrogen Environments: Coupled Stress–Transport–Trapping Mechanisms and Application Gaps

*Elżbieta Gadalińska[1*], Jan Kaufman[1], Jan Šmaus[1], Jan Brajer[1]*

[1]HiLASE Centre, Institute of Physics, Czech Academy of Sciences, Za Radnici 828, 252 41 Dolni Brezany, Czech Republic

*Corresponding author: elzbieta.gadalinska@hilase.cz

**Abstract**

Hydrogen embrittlement limits the deployment of high-strength steels and advanced alloys in hydrogen infrastructure. Laser shock peening (LSP) is increasingly considered as a mitigation route because it combines deep compressive residual stresses with near-surface microstructural modification. This review critically assesses LSP not as an isolated strengthening treatment, but as a surface layer design strategy governed by coupled stress, hydrogen transport and trapping mechanisms. Evidence from steels, nickel based alloys and additively manufactured materials shows that compressive residual stresses may suppress stress assisted hydrogen transport and delay crack initiation, while LSP-induced nanostructuring, dislocations, twins and interfaces can either redistribute hydrogen beneficially or promote localized plasticity and damage. Reported trends are frequently confounded by hydrogen charging mode, surface roughness, contamination, residual stress depth profiling and limited structure performance correlations. Two design critical gaps are identified: the lack of quantitative links between post-LSP stress/defect architectures and hydrogen assisted fatigue crack growth, and the near absence of impact toughness data after LSP under hydrogen exposure. The review proposes mechanism informed qualification routes combining residual stress mapping, hydrogen characterization, service representative mechanical testing and modelling.

## 1. Introduction

Hydrogen embrittlement refers to the degradation of mechanical properties in metallic materials caused by hydrogen ingress, often at concentrations as low as a few parts per million. The phenomenon manifests as reduced ductility, impact toughness, and fatigue resistance, and may ultimately lead to premature, sometimes catastrophic, failures of structural components. Importantly, the severity of degradation can appear disproportionate to the overall hydrogen content and is frequently governed by local conditions, including the stress state, microstructure, and loading history.

As early as the nineteenth century, Johnson demonstrated that hydrogen-induced loss of ductility can be reversible upon hydrogen removal, suggesting that embrittlement arises primarily from hydrogen – metal interactions rather than from permanent microstructural *damage*. This observation remains highly relevant: in many cases, hydrogen does not directly destroy the material, but instead alters its mechanical and microstructural response, shifting deformation and fracture mechanisms. Despite more than a century of intensive research, hydrogen embrittlement remains a complex, multiscale problem spanning phenomena from lattice defects to component-level behaviour and continues to challenge both mechanistic understanding and the development of robust engineering mitigation strategies.

Today, hydrogen embrittlement is among the key factors limiting the deployment of high-strength steels and advanced alloys in hydrogen-related technologies. The problem is particularly critical for components operating in hydrogen-containing environments, for example, high-pressure systems, storage vessels, pipelines, and valves, where even small amounts of hydrogen may trigger significant mechanical degradation, especially under cyclic or dynamic loading and at reduced operating temperatures.

A comprehensive review of hydrogen – metal interactions in gaseous hydrogen environments was provided by Barthélémy [1], covering both fundamental and practical aspects from the distinction between internal and external hydrogen, through hydrogen embrittlement mechanisms, to surveys of engineering failures and design guidance. This perspective emphasises that "material compatibility" in hydrogen service is governed by several coupled factors: hydrogen ingress and transport kinetics, hydrogen trapping in microstructural features, and interactions with stress fields (both residual and applied). Accordingly, the present review adopts this triad: transport, trapping, and stress as its central interpretive framework, as it is essential for assessing the effectiveness of surface engineering strategies aimed at mitigating hydrogen embrittlement.

This motivates a critical assessment of surface engineering approaches that modify both near-surface microstructure and residual stress fields, as these changes can directly influence hydrogen transport, trapping, and damage evolution.

### 1.1. Hydrogen and steels: technological context and embrittlement mechanisms

The contemporary energy system faces the urgent need to reduce $CO_2$ emissions, and hydrogen is increasingly considered a viable energy carrier and fuel. The development of a hydrogen economy offers substantial decarbonization potential, particularly in sectors that are difficult to electrify, in line with global climate commitments including the 2016 Paris Agreement. However, deploying hydrogen technologies requires durable and reliable infrastructure: from high-pressure gaseous hydrogen storage (on the order of hundreds of bar, and up to 700 bar in mobility applications), through transmission pipelines, to valves and other components operating in hydrogen-rich environments.

One of the key material barriers limiting this infrastructure is hydrogen embrittlement (HE). Owing to its small atomic size, hydrogen can enter metals and diffuse through the crystal lattice, leading to a pronounced deterioration of mechanical performance. This degradation may involve reduced ductility, lowered fracture resistance, and accelerated damage accumulation under both static and cyclic loading. As emphasized in reviews on material compatibility in hydrogen service, environmental parameters such as pressure and temperature strongly influence HE and can shift the dominant degradation mechanisms.

Several microscopic mechanisms have been proposed to explain hydrogen embrittlement. The most frequently discussed include Hydrogen-Enhanced Localized Plasticity (HELP), Hydrogen-Enhanced Decohesion (HEDE), and Adsorption-Induced Dislocation Emission (AIDE). In the classical HELP framework, hydrogen facilitates dislocation motion (often rationalized in terms of screening elastic interactions), promoting highly localized plastic deformation in regions of stress concentration, particularly ahead of a crack tip. This concept was established and documented in a seminal manner by Birnbaum and Robertson [2]. In contrast, HEDE described in a thermodynamic formulation, among others, by Oriani and Josephic [3] postulates that hydrogen accumulation in regions of high stress triaxiality (e.g., in front of a crack tip) reduces the local cohesive strength, enabling crack propagation at loads well below the theoretical strength of the material. The AIDE model, discussed in detail by Lynch [4], highlights the role of hydrogen adsorption at crack-tip surfaces and its effect on dislocation emission, which can accelerate crack growth and microvoid coalescence processes.

Importantly, the contemporary understanding of hydrogen embrittlement increasingly departs from treating these mechanisms as mutually exclusive theories. In a more integrated view, HELP and HEDE can coexist, and their relative contribution depends on the local microstructure, hydrogen distribution, and stress state. Robertson and co-workers [1] point out that enhanced dislocation activity (HELP) may drive microstructural evolution and stress concentrations that, in turn, promote conditions conducive to decohesion processes (HEDE). Under gaseous hydrogen exposure, AIDE may further contribute by enhancing localized plasticity and accelerating damage evolution at crack tips [4].

From an engineering perspective, a particularly challenging aspect is that increasing material strength often comes at the expense of greater susceptibility to hydrogen embrittlement. This is especially relevant for high-strength steels, which are structurally attractive (strength-to-weight) yet more prone to abrupt failure scenarios in hydrogen-containing environments. Consequently, there is growing interest in mitigation strategies based on surface engineering, capable of tailoring both residual stress states and the near-surface microstructure. In this context, Laser Shock Peening (LSP) is frequently considered one of the most promising approaches to enhance resistance to hydrogen-assisted degradation, provided that its effectiveness is evaluated in conjunction with hydrogen transport and trapping, as well as the stability of residual stress fields under realistic service conditions.

Against this background, austenitic stainless steels such as SS316L are often regarded as relatively hydrogen-tolerant due to their stable austenitic microstructure, whereas ultra-high-strength martensitic or maraging steels (e.g., Maraging 300) are frequently considered more vulnerable to hydrogen-assisted damage because of their high strength level and defect-rich microstructures. The following section briefly positions these two material classes in the context of hydrogen embrittlement and motivates their selection as representative case studies for assessing the potential and limitations of LSP-based mitigation strategies.

### 1.2. The role of steels: positioning SS316L and Maraging 300 in the context of hydrogen embrittlement

Structural steels play a fundamental role in hydrogen-related infrastructure covering production, storage, and transport owing to their favourable combination of mechanical performance, availability, and technological maturity. Among the material classes considered for hydrogen service, austenitic stainless steels such as SS316L and ultra-high-strength steels, including martensitic and maraging grades (e.g., Maraging 300), are frequently discussed as candidate materials for high-pressure applications. At the same time, these alloys exhibit markedly different susceptibility to hydrogen embrittlement, governed by differences in microstructure, deformation mechanisms, and the kinetics of hydrogen transport and trapping. Understanding these links is essential both for assessing suitability in hydrogen environments and for designing effective mitigation strategies.

SS316L is often regarded as relatively hydrogen-tolerant due to its stable austenitic microstructure; however, it is not immune to hydrogen-assisted degradation. Robertson (2015) [1] emphasized that in austenitic stainless steels the presence of hydrogen can alter fracture behaviour, promoting transgranular cracking with quasi-cleavage features, consistent with hydrogen-induced localization of plastic deformation. This implies that even in alloys commonly considered resistant, local microstructural and stress conditions, and the ability of the microstructure to redistribute or immobilize hydrogen, remain decisive for damage evolution.

Maraging steels, in contrast, offer exceptionally high strength yet may exhibit critical vulnerability to hydrogen-assisted damage, including interfacial decohesion-type processes under certain conditions. Consequently, the use of ultra-high-strength steels in hydrogen service entails elevated risk: Álvarez et al. (2021) [5] reported that even small hydrogen contents can trigger drastic degradation of mechanical properties in this material class. Importantly, they noted that susceptibility to cracking is not governed solely by the ultimate tensile strength, but by the local ductility of the microstructure, which can be sharply reduced in the presence of hydrogen, shifting fracture toward more brittle modes.

Against this background, the role of lattice defects and their organization within the microstructure becomes particularly important. Zhao et al. (2024) [6] compared hydrogen embrittlement resistance in an austenitic steel produced by additive manufacturing (AM) and in its forged counterpart. Despite identical chemical composition, the AM material exhibited higher resistance to hydrogen-assisted cracking, which the authors attributed to a characteristic cellular substructure formed during rapid solidification. Cell walls composed of dense dislocation networks and solute segregation acted as an effective trapping architecture, redistributing hydrogen and limiting its critical localization (e.g., at grain boundaries). This finding supports the broader notion that *defect engineering*, achieved via AM or potentially via surface treatments such as LSP, can mitigate HE by controlling hydrogen trapping and transport pathways.

For high-manganese steels (HMnS), where plastic deformation involves TWIP/TRIP effects, Zhang et al. (2024) [7] highlighted the importance of grain morphology: a homogeneous fine-grained structure more effectively suppresses hydrogen-assisted crack propagation than coarse-grained microstructures. This observation further strengthens the argument that microstructure refinement, including near-surface refinement routes, may be beneficial for certain material classes, provided that the underlying deformation mechanisms are considered.

At the same time, the beneficial effect of AM-induced defect structures is not universal and remains strongly material-dependent. In contrast to austenitic steels, Roudnicka and co-workers [8] reported the opposite trend for Ti-6Al-4V: compared with a conventionally processed counterpart, the PBF-LB/M *as-built* condition showed the highest susceptibility to hydrogen ingress and embrittlement. The authors linked this behaviour to the presence of metastable α′ martensite in the printed microstructure, which exhibits higher hydrogen solubility and cracking susceptibility than the equilibrium α+β microstructure. From the perspective of surface engineering, this implies that for titanium alloys, LSP alone may be insufficient unless combined with appropriate heat treatments that eliminate unstable martensitic phases, an important cautionary insight when translating mitigation concepts across material systems.

These material-specific differences and the central role of microstructure-controlled hydrogen trapping motivate surface engineering strategies, such as laser shock peening, that can deliberately tailor near-surface defect structures and residual stress fields to mitigate hydrogen-assisted degradation.

### 1.3. Mitigation strategies: surface engineering and in-situ approaches

In response to material limitations imposed by hydrogen embrittlement, increasing attention has been directed toward mitigation strategies that modify the near-surface stress state and microstructure. Among these, Laser Shock Peening (LSP) has emerged as a promising surface engineering technique to enhance resistance to hydrogen-induced degradation. This review provides a comprehensive analysis of material responses and experimental methodologies used in studies addressing hydrogen embrittlement and LSP, with particular emphasis on the mechanisms responsible for reported protective effects as well as the potential limitations of the approach.

It is worth noting that LSP is inherently a *post-process* strategy, whereas recent work increasingly highlights the potential of *in-situ* approaches that aim to reduce hydrogen-related defects already during manufacturing. An example is the *on-line laser shielding* concept introduced by Guo and co-workers (2025) [9] for arc-directed energy deposition (Arc-DED). The authors reported that integrating a laser beam with the arc process can enable active suppression of hydrogen-related porosity in real time. The proposed mechanism combines laser ablation (reducing hydrates/contaminants on the wire surface) with improved thermal stabilization of the melt pool, facilitating degassing prior to solidification. As a result, very low residual porosity (<0.05%) was obtained, potentially providing a favourable starting point for subsequent surface treatments. These findings suggest that future mitigation strategies may benefit from a multi-barrier framework that couples defect prevention during manufacturing (e.g., on-line laser shielding [9]) with structural intervention after processing (e.g., LSP), aimed at tailoring residual stresses and near-surface microstructure under hydrogen-relevant service conditions.

The following sections therefore focus on the mechanisms by which hydrogen interacts with microstructure and stress fields, specifically transport kinetics, trapping, and residual-stress stability as these factors ultimately govern the real effectiveness of surface-engineering mitigation strategies.

## 2. Mechanisms Governing Hydrogen Transport and Trapping

### 2.1. Theory of the influence of residual stresses on hydrogen transport

Hydrogen embrittlement (HE) results from complex interactions between hydrogen and metallic microstructure, in which the transport of atomic hydrogen is not governed solely by concentration gradients (Fick's first law) but is strongly coupled to the local stress state. From a thermodynamic perspective, the presence of a stress field modifies the chemical potential of hydrogen; according to the framework discussed by Barthélémy [10], hydrogen migrates from compressive to tensile regions, where the larger lattice volume favours its presence. In this context, Laser Shock Peening (LSP) can act as a critical energetic barrier: the deep compressive residual stresses (CRS) generated by the process locally reduce hydrogen solubility and impose a change in the direction of hydrogen transport, which at the macroscopic scale is reflected in a pronounced reduction of the effective diffusion coefficient ($D_{\mathrm{eff}}$) and a delayed penetration of hydrogen into the bulk [1], [4]**.**

However, contemporary studies addressing the durability of this barrier-employing advanced diffraction methods such as high-energy/high-resolution synchrotron diffraction (HE-XRD) have revealed a dynamic and reversible nature of these interactions. Dabah et al. [11] and Gill et al. [12] demonstrated that hydrogen uptake can lead to local relaxation of CRS. In-situ synchrotron diffraction (EDDI) further enabled observation that the residual-stress state evolves during exposure: stresses often relax during hydrogen charging and then partially recover upon desorption. This behaviour suggests a strong feedback coupling between local hydrogen concentration and the mechanical stability of the near-surface layer.

This correlation is particularly relevant in materials with limited phase stability. While in austenitic steels such as 316L the stability of CRS is generally high, in other grades hydrogen-stress interactions may promote the formation of $\alpha'$martensite. Such a transformation can markedly alter local transport kinetics, potentially turning a protective barrier into a pathway for faster diffusion. The work by Örnek et al. [13], combining correlative synchrotron analysis with digital image correlation (DIC), provided evidence that hydrogen infusion induces local tensile stresses within the austenitic phase, which leads to localized plastic deformation (HELP) and crack nucleation at interphase boundaries.

Understanding these mechanisms enables more precise design of LSP processes. Niwa et al. [14], using finite element modelling and spherical indentation tests, showed that circumferential tensile stresses at the boundary of the treated zone are responsible for crack initiation, whereas the interior of the CRS zone retains full integrity. This conclusion, together with observations by Tien et al. [15] regarding defect reorganization, indicates that the effectiveness of HE mitigation by LSP depends not only on the initial CRS magnitude, but also on the resistance of the induced microstructure to hydrogen-assisted redistribution of stresses and lattice defects.

Modern approaches to optimizing LSP for components with complex geometry rely on advanced numerical modelling, often using the Johnson–Cook constitutive equation. Alinaghian et al. [16] demonstrated that correlating finite element simulations with experimental residual-stress measurements (e.g., the slitting method) allows critical regions to be identified with high precision, such as the edges of technological holes where stress concentration is highest. Applying LSP in specific triangular regions around such stress concentrators can reduce the resultant tensile stress field, directly delaying the initiation of fatigue and hydrogen-assisted cracking. Fundamental analysis by Robertson [1] provides further insight into the role of stress fields. It was shown that hydrogen not only migrates toward regions of high triaxial tensile stress, but through interactions with dislocation stress fields can actively redistribute local stress states within the material. Introducing stable CRS fields via LSP therefore not only creates a diffusion barrier, but may also stabilize the microstructure by limiting hydrogen-assisted mobility of defects, thereby inhibiting crack development at the atomic scale.

The description of hydrogen transport in plastically deformed materials is commonly based on the assumption of local thermodynamic equilibrium between lattice hydrogen and hydrogen trapped at defects, known as Oriani's equilibrium. Within this framework, the population of hydrogen in traps (such as dislocations generated by LSP) is directly linked to the chemical potential of lattice hydrogen. In the context of LSP, this implies that the barrier stability of the near-surface layer depends not only on the depth of compressive stresses, but also on the ability of the introduced defects to bind hydrogen in a sustained manner under thermodynamic equilibrium conditions [3].

Recent synchrotron work reported by Lightsources.org [17] relating to studies by the Örnek and Pan team highlights the importance of surface interactions. Using grazing-incidence X-ray diffraction (GIXRD) and in-situ X-ray reflectometry (XRR), the authors reported that degradation of stainless steel can initiate with chemical reduction of the passive film. Hydrogen not only diffuses through oxides, but actively interacts with the passive layer, generating point defects (oxygen vacancies) and causing thinning and local loss of stability. This insight is highly relevant for LSP: it suggests that laser-induced surface nanostructuring must be optimized not only to introduce beneficial stresses, but also to promote rapid repassivation and the formation of a denser, defect-resistant oxide layer capable of resisting hydrogen-driven reduction.

Recent review work, including Yu et al. [18], proposes extending classical HELP/HEDE descriptions by introducing the *defactant* concept. According to this idea, hydrogen atoms behave as *defectants* (analogous to surfactants), reducing the formation energy of lattice defects such as vacancies or dislocation loops. In the context of LSP, this implies that in the presence of hydrogen, nanostructure generation during peening may occur at lower energetic cost, while the stability of the resulting structure under service conditions may be challenged by the thermodynamic tendency of hydrogen to segregate into newly formed defects.

A particularly useful perspective on the hierarchy of factors governing hydrogen-assisted degradation is provided by Shibayama et al. [19]. In U-bend tests of an ultra-high-strength martensitic steel (~1.5 GPa), the authors decoupled the effects of plastic strain from residual stresses. They showed that the dominant criterion for crack initiation is the peak maximum principal stress, rather than the local defect density associated with plastic deformation. The apparent increase in HE susceptibility with increasing plastic strain was in fact driven by the evolution of the residual-stress gradient, rather than by hydrogen trapping at dislocations alone. This provides strong support for the rationale behind LSP: in steels of such high strength, precise control of the stress field (which LSP offers) may be the key lever for HE mitigation, while microstructural changes can be of secondary importance.

Taken together, these studies highlight a dynamic coupling between hydrogen, stress fields, and microstructural evolution, which is central to assessing LSP-based mitigation concepts. Yet, stress-assisted transport alone cannot explain the large scatter in hydrogen embrittlement responses reported across materials and processing states. A critical missing piece is hydrogen trapping, which governs the partitioning of hydrogen between lattice and trap sites (e.g., under Oriani-type equilibrium) and thereby controls the effective diffusion behaviour as well as the local hydrogen concentration available for damage processes.

### 2.2. Hydrogen Trapping

A range of mechanisms has been proposed to explain hydrogen embrittlement, including Hydrogen-Enhanced Localized Plasticity (HELP), Hydrogen-Enhanced Decohesion (HEDE), Adsorption-Induced Dislocation Emission (AIDE), and hydrogen-stabilized vacancy formation. Each mechanism emphasises a different facet of hydrogen-microstructure interactions, spanning from enhanced dislocation mobility to reduced cohesive strength at interfaces such as grain boundaries and phase boundaries. Importantly, these mechanisms should not be regarded as mutually exclusive. Their relative contribution to mechanical degradation depends strongly on the local microstructure, stress state, and hydrogen distribution. Consequently, the dominant embrittlement mechanism may vary not only across different materials but also across different regions within the same component, particularly in the presence of complex residual stress fields and heterogeneous microstructures.

Multiphase microstructures play a major role in both hydrogen transport and trapping. In materials containing ferritic and austenitic phases, interphase boundaries can act as diffusion barriers, promoting local hydrogen accumulation and increasing the likelihood of crack initiation. Experimental observations further indicate a dynamic coupling between mechanical stress fields and microstructural features responsible for hydrogen trapping, which complicates both description and prediction of material behaviour under hydrogen exposure.

Hydrogen trapping within metallic microstructures is therefore a key factor governing susceptibility to hydrogen embrittlement. Depending on the nature of microstructural defects, traps are commonly classified as reversible or irreversible. Reversible traps include primarily dislocations, grain boundaries, subgrain boundaries, and deformation twins, where hydrogen can be temporarily bound and relatively readily released in response to changes in temperature, stress, or environmental conditions. These traps significantly affect transport kinetics by modifying the effective diffusion behaviour and by shaping local hydrogen availability in mechanically loaded regions. In contrast, irreversible traps, such as non-metallic inclusions, voids, cavities, and certain volumetric defects, exhibit substantially higher binding energies. Their presence leads to persistent local hydrogen accumulation, which can promote microcrack initiation and accelerate mechanical degradation, particularly under impact or cyclic loading.

In the context of surface treatments such as peening-based processes, including Laser Shock Peening (LSP), the character of the generated hydrogen traps requires critical assessment. On the one hand, intense plastic deformation increases dislocation density and refines the microstructure, which can increase the population of reversible traps and potentially delay hydrogen transport into the bulk. On the other hand, overly severe processing can generate volumetric defects, such as voids or local discontinuities, that act as irreversible traps and can favour hydrogen accumulation at critical sites. Accordingly, the effectiveness of peening-based approaches in mitigating hydrogen embrittlement should not be evaluated solely by the magnitude and depth of compressive residual stresses, but also by the nature, stability, and distribution of the hydrogen traps created. Distinguishing between *beneficial* and potentially *detrimental* trap populations is essential for interpreting the scatter in reported literature outcomes and for optimising surface-treatment parameters for hydrogen service.

Lynch [4] noted that experimentally distinguishing between HELP and AIDE is often challenging. Nevertheless, AIDE is particularly effective in rationalising cracking scenarios in which hydrogen is supplied from a gaseous environment and transported to the crack tip (e.g., in pressurised $H_2$ vessels), whereas HELP tends to dominate in cases where hydrogen is primarily dissolved within the bulk. For high-strength steels, Lynch suggested that degradation often reflects a combination of AIDE (near the surface/crack tip) and HEDE (deeper in the material, for instance along grain boundaries).

#### 2.2.1. Trap classification by binding energy and implications for LSP-induced microstructures

According to the recent review by Chen and co-workers [20], the reversible/irreversible classification can be formalised via the energetic barrier for desorption. At room temperature, traps with binding energies $|E_b| > 50$ kJ/mol (e.g., incoherent TiC carbides or oxides) are often treated as irreversible, immobilising hydrogen and limiting its migration to critical sites. In contrast, defects commonly generated by LSP, such as grain boundaries ($\sim$ 9–49 kJ/mol) and dislocation stress fields ($\sim$ 12–27 kJ/mol), are typically classified as reversible traps. Their role is inherently dual: they can slow the effective diffusion ($D_{\mathrm{eff}}$) by temporarily storing hydrogen, yet under changes in loading conditions they may also release hydrogen and thereby feed processes consistent with HELP. A key conclusion emerging from Chen's analysis is that highly organised defect architectures, such as dislocation cell walls, may remain beneficial even at high defect densities by effectively redistributing hydrogen and preventing critical localisation.

#### 2.2.2. Grain boundary engineering (GBE) and twin boundaries as *hydrogen-resilient* interfaces

Modern approaches to designing hydrogen-resistant materials increasingly move beyond treating grain boundaries solely as weak links and instead adopt the concept of Grain Boundary Engineering (GBE). As highlighted by Li et al. [21], mitigating embrittlement requires controlling boundary character, not only boundary density. High-angle grain boundaries (HAGBs) with high interfacial energy often provide preferential diffusion pathways and segregation sites for hydrogen, promoting decohesion-type behaviour (HEDE). By contrast, special low-energy boundaries, such as low-Σ coincidence site lattice (CSL) boundaries, exhibit substantially higher resistance to intergranular cracking. Processing strategies (potentially including LSP, depending on material and parameter regime) that increase the fraction of special boundaries may therefore strengthen the interfacial network, creating barriers to crack propagation without necessarily compromising strength.

Twin boundaries deserve particular attention in this context, especially given that LSP can generate a high density of deformation twins in suitable alloys. Li et al [21] emphasise the distinction between coherent and incoherent twin boundaries. Coherent twin boundaries (CTBs) can act as effective obstacles to dislocation motion (strengthening) while exhibiting a low propensity for hydrogen uptake, making them comparatively safe interfaces. A dense network of nanotwins, if introduced in a controlled manner, may therefore provide an attractive synergy: higher hardness/strength combined with improved resistance to hydrogen-assisted cracking, as CTBs are less prone to crack initiation than typical high-energy grain boundaries [21]. This perspective supports the broader view that plastic deformation induced by LSP can be beneficial if the resulting interface population and defect topology are properly engineered.

#### 2.2.3. Nano-precipitates as ideal irreversible traps: trap engineering

A critical component of hydrogen-resistant design is trap engineering, i.e., the deliberate introduction of irreversible traps through nanoscale precipitates. According to Li et al. [21], fine incoherent precipitates (e.g., vanadium or titanium carbides in steels) can act as *reservoirs* that sequester hydrogen. With high binding energies (often $E_b > 60$ kJ/mol), such precipitates immobilise hydrogen and prevent its diffusion to stress concentrators (e.g., a crack tip). In contrast to reversible traps (such as dislocations) that may release hydrogen under load and thereby sustain HELP-type processes, irreversible precipitate traps can reduce the concentration of mobile lattice hydrogen and stabilise the material even under severe service conditions [21].

#### 2.2.4. Direct evidence relevant to HELP and defect evolution under hydrogen

One of the most challenging aspects of HELP to verify experimentally is direct observation of defect evolution within the material. Cho et al. [22], using in-situ neutron diffraction and diffraction line profile analysis (DLPA), provided quantitative evidence for hydrogen-induced changes in defect evolution in an austenitic high-manganese steel. They reported that in hydrogen-charged specimens the dislocation density increased more slowly during the early stages of deformation compared with hydrogen-free specimens, consistent with enhanced dislocation mobility and annihilation (in line with HELP), and with delayed work hardening. For LSP-relevant microstructures, this suggests that the dislocation-based barrier introduced by peening may undergo dynamic rearrangement and potentially partial weakening under hydrogen exposure more rapidly than predicted by purely static models.

Chen et al. [20] also emphasise that trap classification based on binding energy provides a mechanistic link between microstructure and effective transport. Reversible traps with lower binding energies such as dislocation stress fields ($\sim$ 12–27 kJ/mol) and certain coherent twin boundaries may act as temporary hydrogen stores, but can also become hydrogen sources under applied stress. By contrast, irreversible traps such as incoherent TiC ($E_b \approx 60$–$129$ kJ/mol) or vanadium carbides (e.g., $V_4C_3$) immobilise hydrogen and reduce diffusivity. On this basis, the authors advocate *trap engineering* through dense populations of fine incoherent precipitates that act as hydrogen *fuses*, capturing hydrogen before it reaches critical stress-concentration zones.

#### 2.2.5. Vacancies and HESIV – additional mechanism relevant to severe plastic deformation

Beyond HELP and HEDE, Chen et al. [20] highlight the role of Hydrogen-Enhanced Strain-Induced Vacancies (HESIV). Hydrogen can stabilise deformation-generated vacancies, promoting their agglomeration and the formation of nanoscale voids. This mechanism is particularly relevant for processes involving large plastic strains, including LSP. If the process generates excessive vacancy populations in the near-surface layer, hydrogen may accelerate their coalescence into microcracks even in the absence of substantial external loading. This underscores the need to optimise LSP parameters not only with respect to residual stresses and dislocation structures, but also to minimise point-defect populations that can become damage nucleation sites under hydrogen exposure.

#### 2.2.6. Diffusion tortuosity in nanostructured layers

Maurya and Akhtar [23] introduced the concept of diffusion tortuosity to describe the barrier performance of nanostructured layers produced, among others, by severe surface treatments. In a nanograined structure, hydrogen does not diffuse along a straight path but is forced to navigate an extremely dense and tortuous network of grain boundaries. The authors argue that this geometric lengthening of diffusion pathways, combined with trapping at triple junctions, can strongly reduce the effective diffusion coefficient ($D_{\mathrm{eff}}$). For LSP, this implies that even a relatively thin strengthened layer (on the order of hundreds of micrometres) may act as an effective *labyrinth*, delaying hydrogen ingress to the bulk by orders of magnitude more than would be expected from layer thickness alone.

# 3. Surface engineering

## 3.1. Laser Shock Peening (LSP): a comprehensive analysis

Mechanical surface treatment methods introduce deep compressive residual stress fields that can strongly influence crack initiation and crack growth. This effect is widely regarded as beneficial in terms of improving fatigue performance and resistance to environmentally assisted cracking. However, the presence of hydrogen may modify the stability of compressive residual stresses, particularly in near-surface regions where hydrogen concentration gradients are typically the highest.

Laser Shock Peening (LSP), compared with conventional peening methods such as shot peening, is distinguished by its ability to generate substantially deeper compressive residual stresses. The LSP process intensely modifies the near-surface layer by introducing a high dislocation density, refining grains down to the nanostructural range, and promoting the formation of mechanical twins. As a result, increased surface hardness and strength are obtained, which many studies associate with improved resistance to hydrogen-induced degradation. From the perspective of hydrogen–material interactions, the effectiveness of LSP derives from a combination of two coupled effects. First, deep compressive residual stresses reduce the local chemical potential of hydrogen, thereby limiting hydrogen transport into the bulk. Second, microstructural modification produces a high density of crystalline defects that act as hydrogen traps. This combination of residual stress engineering and deliberately altered microstructural features makes LSP a promising mitigation tool for hydrogen embrittlement. At the same time, these mechanisms are not unambiguously beneficial under all conditions. While compressive residual stresses can effectively suppress hydrogen-assisted cracking, the associated microstructural refinement and increased defect density may promote enhanced hydrogen trapping and localized plasticity. Consequently, LSP may simultaneously reduce hydrogen transport and activate competing embrittlement mechanisms, particularly in materials with complex microstructures and under impact or cyclic loading.

In austenitic steels such as AISI 316L, LSP leads to grain refinement and a high dislocation density, which constitute the primary strengthening mechanism and may contribute to improved resistance against hydrogen-induced cracking. Nevertheless, given the documented sensitivity of residual stress fields to hydrogen exposure, the assumption that LSP-induced stresses remain unchanged in hydrogen-rich environments requires critical verification. Although classical review papers do not always address Laser Shock Peening explicitly, observations regarding surface condition, stress fields, and hydrogen trapping allow a logical extension toward surface engineering technologies. Interactions between mechanical stress fields and hydrogen behaviour provide a robust basis for arguing the role of surface treatments as mitigation tools for hydrogen embrittlement.

#### 3.1.1. Microstructural refinement and stress-depth effects

One of the key outcomes of Laser Shock Peening is intensive microstructural modification of the near-surface layer. The process leads to a substantial increase in dislocation density, formation of subgrains, and, in many cases, refinement toward a nanocrystalline structure. In materials with low stacking fault energy (SFE), such as many austenitic steels, deformation twinning is additionally observed, further increasing microstructural complexity and influencing plastic deformation mechanisms. The modified near-surface layer exhibits increased hardness and enhanced resistance to crack initiation, directly linked to the fine-grained structure and the high density of crystalline defects.

In parallel with microstructural modification, LSP generates deep compressive residual stresses whose affected depth significantly exceeds that typically obtained by conventional shot peening. Depending on processing parameters and material type, the depth of compressive stress influence may range from several hundred micrometres to several millimetres. Such deeply embedded stress fields play an important role in suppressing crack initiation and propagation and in modifying hydrogen transport by lowering the local chemical potential of hydrogen in compressive regions.

Available literature data indicate that after LSP a significant reduction in the effective hydrogen diffusion coefficient, $D_{\text{eff}}$, is observed in various steel types, including austenitic steels, duplex steels, and ultra-high-strength steels [24], [25], [26]. The reduction in $D_{\text{eff}}$ is attributed both to the presence of compressive residual stresses and to enhanced hydrogen trapping in LSP-induced microstructural defects [24], [27]. Although absolute values of $D_{\text{eff}}$ reduction depend on the material and the measurement approach, the trend of slowed hydrogen transport after LSP is reported consistently across multiple studies [24], [25], [28].

It should be emphasized, however, that the vast majority of available data on the influence of LSP on $D_{\text{eff}}$ and residual-stress stability have been obtained under laboratory conditions, using specimens of limited thickness and simplified geometry. Translating these findings to real thick-section components is not trivial and requires accounting for additional factors such as stress-field heterogeneity, microstructural gradients, and complex service loading. Scaling LSP effects from laboratory coupons to engineering components therefore remains a significant challenge and one of the key areas requiring further investigation.

#### 3.1.2. Evidence in 316L: SSRT performance and parameter *process window*

The effectiveness of LSP in mitigating hydrogen embrittlement effects in 316L steel was confirmed in the work of Agyenim-Boateng et al. [29], where a pronounced reduction in ductility loss was observed. In SSRT tests, laser-treated samples showed an elongation loss of only 9.66%, compared with nearly 14% for as-received specimens. The authors attributed this effect to the synergistic action of deep compressive residual stress fields and surface nanostructuring, which by increasing the number of trapping sites (dislocations, grain boundaries) effectively controls hydrogen absorption and transport.

For austenitic 316L, LSP thus demonstrates a significant capacity to mitigate HE. SSRT results indicate that LSP can reduce elongation loss from 13.93% (as-received) to 9.66%. This mechanism is further supported by an increase in surface hardness by more than 10% and the introduction of a high density of dislocations and subgrain boundaries, which redistribute hydrogen and prevent its critical accumulation [16].

The effectiveness of the hydrogen barrier generated by LSP is strongly correlated with laser power density; however, this relationship is not monotonically beneficial. As shown by Attolico et al. (2022) [30], increasing power density leads to a linear increase in maximum compressive residual stresses and their depth of influence (remaining stable down to ~0.8 mm). Nevertheless, the authors caution against uncontrolled increases in processing energy. Their results show that aggressive processing parameters can rapidly degrade surface topography, an increase in roughness parameters ($R_a$, $R_t$) becomes a critical factor that may offset the benefits of compressive stresses by creating micro-notches that initiate cracking. In the context of HE mitigation, this implies that for each material there exists a *processing window* in which the depth of the diffusion barrier (the CRS zone) is maximized while surface integrity is preserved, thereby preventing activation of AIDE.

#### 3.1.3. Beyond steels: nickel-based superalloys and microstructural *transport highways*

Extending LSP to nickel-based superalloys has provided new mechanistic insights into HE mitigation. In a recent study by Sridharan et al. [31] published in *Optics & Laser Technology*, the authors showed that for Inconel 718 an important protective role is linked to carbide modification. They identified a distinctive mechanism: LSP fragments carbides at grain boundaries and generates a dense dislocation network that counteracts the formation of so-called *diffusion highways*. In the untreated condition, hydrogen-assisted vacancy formation at grain boundaries promotes rapid hydrogen migration and intergranular cracking. LSP, by introducing deep compressive stresses and altering precipitate morphology, blocks these fast transport paths and forces hydrogen to be trapped in safer intra-granular locations. This supports the view that LSP may be effective not only for steels but also for critical hydrogen-exposed components in aerospace and energy applications.

A recent analysis by Wang et al. [28] further clarified the correlation between LSP-induced microstructure and hydrogen damage sensitivity. The authors reported that a key effect of LSP in 316L is not only dislocation generation but also a pronounced increase in the fraction of low-angle grain boundaries (LAGBs) and the formation of stacking faults. In electrochemical charging tests with variable current density, a fundamental difference in surface degradation mechanisms was observed: while untreated samples exhibited severe damage with deep pits and *holes* (interpreted as traces of ruptured hydrogen blisters), the LSP-treated surface maintained substantially higher integrity, with damage limited to shallow microcracks. This behaviour was attributed to LAGBs and stacking faults acting as dispersive traps that homogenize hydrogen distribution in the near-surface layer, preventing local accumulation to critical pressures capable of rupturing the material from within.

#### 3.1.4. Process-energy selection, precipitation-hardened analogies, and *defect architecture*

The question of pulse energy selection remains non-trivial and strongly material-dependent. In contrast to aluminium alloys, where excessive energy degrades surface quality, aggressive parameters may be beneficial for ultra-high-strength materials. Recent work by Li et al. [21] on a Cu–Ti–Fe alloy contributes to this discussion. Although the studied material is a copper alloy, its precipitation-hardening mechanism shows striking similarities to the metallurgy of maraging steels. The authors reported a direct correlation: higher laser power density led to more intense near-surface grain refinement. Notably, it was the surface nanostructuring (not only the stresses) that was identified as the decisive factor in reducing hydrogen embrittlement. A dense grain-boundary network in the nanocrystalline layer acted as an effective barrier to hydrogen diffusion. This suggests that for precipitation-hardened alloys such as maraging steels, LSP strategies may benefit from targeting maximized grain refinement through higher energy densities, even at the cost of a modest increase in roughness.

The apparent paradox that LSP introduces lattice defects that might theoretically increase HE susceptibility is addressed by Chen et al. [32]. They argued that while individual dislocations can facilitate hydrogen transport, heavily deformed structures that form dislocation cells act as a dense trap network that locks hydrogen within cells, limiting its availability at crack tips. This implies that LSP-induced surface nanostructuring that produces such a defect architecture can be desirable.

#### 3.1.5. Multiphase microstructures, nanocarbidization, and strong trapping architectures

The influence of LSP on the microstructure of multiphase alloys can extend well beyond classical dislocation-based hardening. In the case of cast nickel-based superalloys such as MAR-M247, Zhang et al. [33] reported that the shock wave can induce the formation and redistribution of nanocarbides within the metallic matrix. Coupled experimental characterization and finite element modelling (FEM) indicated that high laser power density generates extremely high dislocation densities, which act as nucleation sites for fine carbide precipitates. These nanocarbides may provide a dual protective function in the context of hydrogen embrittlement: (i) they mechanically impede dislocation motion, increasing hardness, and (ii) within a trapping framework, they can act as strong, irreversible hydrogen traps (high binding energy), immobilizing hydrogen and limiting its migration toward critical grain-boundary regions.

Strong support for the pursuit of near-surface nanostructuring via LSP also emerges from recent results on high-manganese steels. Zhang et al. [7] demonstrated that precise control of grain size and grain-size distribution constitutes an effective strategy for mitigating HE. In in-situ electrochemical charging experiments, increasing grain boundary density (via grain refinement) produced a *hydrogen dilution effect*: the dense boundary network behaves as a distributed trap system, reducing the local hydrogen concentration per unit boundary area and preventing the attainment of critical conditions for intergranular cracking. For LSP, this implies that the generated nanocrystalline layer is not only a mechanical barrier (through hardening), but also a capacity-type trapping architecture that stores hydrogen away from the active fracture process zone.

#### 3.1.6. Welded joints as critical weak links and post-weld LSP

Welded joints remain a critical vulnerability in hydrogen storage and transport infrastructure, as the thermal nature of welding processes creates localized regions prone to environmental degradation. In their recent review, Roshith and Jose [34] emphasized that conventional welding techniques (TIG, MIG, and even laser beam welding - LBW) inevitably introduce detrimental tensile residual stresses (TRS), concentrated in the fusion zone and heat-affected zone (HAZ). In hydrogen environments, these tensile regions can act as *diffusion pumps*, accelerating hydrogen migration into the material and promoting crack initiation, including stress corrosion cracking (SCC).

Applying LSP as a post-weld treatment can effectively invert this stress state. Roshith and Jose [34] reported that LSP transforms tensile stresses into deep compressive residual stresses (CRS), covering not only the weld region but also the critical HAZ. A characteristic outcome is a W-shaped microhardness profile, in which local hardness peaks in the weld and base material regions reflect intensive grain refinement. The authors described a four-stage microstructural evolution: generation of dislocation lines, formation of dense dislocation walls (DDWs), transformation into subgrain boundaries, and finally recrystallization toward nanometric grains. From the HE-mitigation perspective, this evolution is ambivalent but predominantly beneficial: nanostructured grain boundaries provide a dense trap network that, combined with compressive stresses, suppresses fatigue and corrosion-assisted crack propagation. An additional advantage of LSP relative to conventional shot peening is the absence of surface contamination and the typically lower surface roughness, which reduces the likelihood of notch-like surface features that could initiate hydrogen-assisted cracking on pipelines or vessels.

#### 3.1.7. Additively manufactured stainless steels: anisotropy, SCC, and local stress homogenization

In contrast to conventionally processed materials, additively manufactured (AM) stainless steels often exhibit pronounced anisotropy due to the directional nature of layer-by-layer building. Recent work by Over and Yao [35] provided new insight into the role of LSP in mitigating stress corrosion cracking (SCC) in such microstructures. In U-bend corrosion tests (boiling $MgCl_2$solution), LSP significantly delayed crack initiation relative to the as-built condition. Importantly, the effectiveness of the treatment depended on build orientation: horizontally built (XY plane) specimens exhibited longer life in the corrosive environment than vertically built (XZ) specimens. The authors suggested that LSP not only introduces a global compressive residual stress field but also modifies local back-stress, reducing the hardening anisotropy typical of AM microstructures. This implies a dual role of LSP in AM materials: (i) a barrier effect through compressive stresses and (ii) a homogenization of mechanical response, highly relevant to the design of durable hydrogen infrastructure components with complex geometries.

#### 3.1.8. Mechanistic dependence on crystal structure and stacking-fault energy (SFE)

The nanostructuring response to shock-wave loading is strongly dependent on crystal structure and stacking-fault energy (SFE). As summarized by Jia et al. (2024) [36], in face-centered cubic (FCC) materials with low SFE, the dominant mechanism is mechanical twinning (MTs). Multiple laser impacts can generate twins in multiple orientations, subdividing grains into sub-micron blocks that may ultimately undergo dynamic recrystallization. Conversely, in high-SFE materials (e.g., aluminium) and in body-centered cubic (BCC) structures (e.g., ferritic steels), dislocation slip dominates: the shock wave produces dense tangles and dislocation walls that evolve into subgrain boundaries. Understanding these differences is critical for designing hydrogen barriers: in FCC structures, a dense network of coherent twins can constitute a stable and effective trapping architecture, whereas in BCC structures careful control of dislocation density is needed to avoid excessive hydrogen-assisted dislocation mobility (HELP) that could undermine the intended strengthening and barrier effects.

#### 3.1.9. Residual-stress holes (RSH) and the importance of stress-field uniformity

A technologically important hazard that may influence local susceptibility to corrosion and hydrogen ingress is the so-called residual stress hole (RSH) phenomenon. Jia et al. [36] noted that for conventional Gaussian circular laser spots, edge effects and the convergence of rarefaction waves toward the spot center can lead to a localized reduction of compressive stress at the geometric center of impact. Under hydrogen exposure, such a locally reduced compressive region may become a weak link, a local diffusion window. A practical mitigation route is the use of flat-top beam profiles and square spots, which have been reported to suppress the RSH effect and to provide a more uniform compressive stress field and greater depth of influence (reported as up to ~36% greater compared with circular spots).

A closely related set of findings was reported by Chen et al. [20] for 20CrNiMo steel, where the authors identified RSH as a dangerous local drop in compressive residual stress at the spot center. The phenomenon was linked to shock-wave interference and correlated strongly with overlap rate. At low coverage, these local *holes* became the weakest locations, drastically reducing fatigue life despite a high average compressive stress level. In the context of HE mitigation, any discontinuity in a compressive stress field represents a potential hydrogen entry/transport pathway, motivating stringent overlap strategies (e.g., overlap > 60%) to ensure the integrity of the stress-based barrier. More broadly, these results suggest a shift in design philosophy from *high power* toward *high uniformity*. Chen et al. [20] argued that improved stress-field uniformity achieved through higher overlap can compensate for lower laser energy density in fatigue performance. For hydrogen-service components, this offers a route to milder processing parameters that reduce the risk of surface roughness degradation and microcrack formation (consistent with observations by Dyer et al. [37]) while maintaining high fatigue performance through continuity of the compressive layer. Such a strategy also reduces the likelihood of activating AIDE by minimizing surface notches without sacrificing the residual-stress benefit.

### 3.1.10. Surface topography: dual response and scatter reduction in AM components

The effect of LSP on surface topography is non-unique and strongly dependent on the initial surface condition, which is an issue of critical importance for hydrogen applications sensitive to notch effects. Dyer et al. [37] reported a dual surface response in additively manufactured Ti-6Al-4V (L-PBF). For rough as-built surfaces, LSP applied without a sacrificial layer was beneficial: local melting of loosely attached powder particles and blunting of sharp valleys reduced the likelihood of crack initiation. In contrast, for pre-machined smooth surfaces, the same LSP approach degraded surface quality by introducing new irregularities and micro-indentations. This is an important design conclusion: for high-smoothness hydrogen components, LSP without a sacrificial/ablative layer may be risky, as induced roughness can offset the benefits of compressive stresses by creating new hydrogen adsorption sites.

From a deployment perspective, repeatability is also critical for AM components in energy systems. AM materials exhibit intrinsically large scatter in fatigue properties due to the stochastic nature of defects (pores, lack-of-fusion). Dyer et al. [37] observed that while LSP did not always dramatically increase the mean fatigue life of machined-surface specimens, it significantly reduced the scatter. This effect can be rationalized by homogenization of the near-surface stress state: deep compressive stresses mask the influence of small, random subsurface defects, making the mechanical response more predictable. For critical infrastructure (e.g., $H_2$ pipelines), such stabilization of performance may be as valuable as an increase in nominal strength.

### 3.1.11. Quantitative evidence in AM 316L and hydrogen-embrittlement index

Recent 2025 studies provide quantitative evidence for the effectiveness of LSP in high-energy-density additively manufactured materials (L-DED). Ha [38] reported that in L-DED 316L steel, multiple LSP passes (sevenfold LSP) led to a pronounced reduction in a hydrogen-embrittlement index (HE index). In SSRT tests, the reduction in elongation decreased from 36.1% (untreated condition) to only 12.5% after LSP. The authors attributed this result to a synergy of three factors: (i) removal of tensile residual stresses typical of DED, (ii) deep compressive residual stress introduction, and (iii) grain refinement increasing hydrogen-trap density, thereby dispersing hydrogen in the near-surface layer and delaying crack nucleation.

### 3.1.12. Fundamental microstructural mechanisms: dislocations, twinning, and gradient nanostructures

A fundamental strengthening mechanism in LSP, described in detail by Cao et al. [39], is the synergy between dislocation generation and mechanical twinning. The shock wave induces severe plastic deformation, increasing dislocation density by orders of magnitude. These dislocations entangle and form dense dislocation walls and dislocation cells, which act as effective barriers to subsequent defect motion (Taylor hardening). In parallel, in materials with lower SFE, grains are fragmented by twinning. In the context of HE mitigation, the resulting microstructure plays a dual role: it increases surface hardness (hindering crack initiation) and creates a dense hydrogen-trap network, which if stable redistributes hydrogen within the strengthened layer. A key outcome is near-surface nanocrystallization, which contributes to improved mechanical properties consistent with Hall–Petch strengthening. Cao et al. [39] emphasized that LSP can refine grains from the micrometre to the nanometre scale. This strong increase in grain-boundary area not only raises hardness and yield strength but also alters environmental interaction kinetics. In hydrogen environments, the nanostructured layer can act as a diffusion filter: a high density of grain boundaries reduces effective hydrogen diffusion into the bulk, provided that boundaries are not weakened by deleterious segregation.

In a recent review of surface-modification strategies, Maurya and Akhtar (2026) [23] defined a key advantage of LSP as the ability to produce a gradient nanostructured surface (GNS) layer. Unlike coatings, which introduce a sharp interface prone to delamination under hydrogen pressure, LSP produces a smooth transition from a nanocrystalline surface to an ultrafine-grained (UFG) zone and then to the bulk material. Such architecture is highly relevant to HE mitigation: the nanostructured surface provides a diffusion barrier (high grain-boundary density), while the gradually changing underlying structure reduces stress localization that could otherwise promote subsurface cracking. Focusing on transport kinetics, Maurya and Akhtar emphasized that surface modification by SMAT and LSP can significantly reduce hydrogen permeability. The mechanism is not solely trapping, but also diffusion-path tortuosity within nanocrystalline structures. Compared with conventional methods, LSP offers a distinct advantage: it can deliver this effect without introducing chemical contaminants or surface microcracks, which are common side effects of aggressive shot peening that may become crack nucleation sites in hydrogen atmospheres. A further, often underappreciated aspect discussed by Maurya and Akhtar is the effect of nanostructuring on passive-film quality. Evidence summarized in the review suggests that on LSP-treated surfaces (especially austenitic steels and aluminium alloys) the oxide film can become denser and less defective compared with coarse-grained material. This is attributed to the increased surface grain-boundary density, which provides nucleation sites for rapid oxide formation and repassivation. Such an improved oxide layer can serve as a first physicochemical defence line, reducing hydrogen adsorption (Volmer step) before atomic hydrogen enters the lattice.

Finally, Huang et al. [24] highlighted the ambivalent role of the dislocation density introduced by LSP. While dislocation pile-ups increase hardness and strength, in hydrogen environments they may also become precursors to cracking. The authors discussed evidence that excessively high dislocation densities, beyond a saturation threshold, may promote local hydrogen supersaturation and activation of HELP, leading to premature strain localization. Accordingly, they advocated optimizing LSP toward a *balanced microstructure*, in which stable boundary-type traps dominate rather than loosely tangled dislocation networks that can facilitate hydrogen transport. An additional point raised in Huang's review concerns passive-film stability: unlike contact-based methods (e.g., burnishing), LSP does not necessarily disrupt oxide continuity and may even enhance it. XPS evidence cited by the authors suggests an increased fraction of more stable oxides (e.g., $Cr_2O_3$ on stainless steels or $TiO_2$ on titanium alloys), which can act synergistically with the subsurface stress barrier by limiting hydrogen adsorption and dissociation at the surface ($H_2 \rightarrow 2H$).

### 3.2. Comparative Analyses of Impact-Based Surface Treatments

One of the key advantages of Laser Shock Peening (LSP) over conventional impact-based surface treatments such as shot peening is the minimal degree of surface damage. LSP can yield relatively low surface roughness (low $R_a$) and eliminates the risk of introducing foreign contaminants into the near-surface layer. These aspects are particularly relevant for components intended to operate in high-purity hydrogen environments, where surface contamination may promote local hydrogen accumulation and trigger degradation. Another important benefit of LSP is the ability to precisely control processing parameters, including laser power density, spot size, and surface coverage. This enables deliberate design of the near-surface condition and optimization of both microstructural stability and compressive residual stresses under hydrogen exposure. In contrast to conventional methods, where the extent of microstructural modification and residual stress generation can be difficult to control unambiguously, LSP offers improved repeatability and flexibility in tailoring treatment outcomes to specific application requirements.

At the same time, it should be emphasized that the effectiveness of LSP in mitigating hydrogen embrittlement cannot be attributed solely to the presence of compressive residual stresses. While compressive stresses can suppress hydrogen-assisted cracking by reducing crack opening and limiting stress-assisted hydrogen transport, the accompanying microstructural modification introduces competing effects. Increased dislocation density, refinement of grains or subgrains, and a higher density of lattice defects can promote hydrogen trapping and localized plasticity. As a result, LSP may simultaneously mitigate and exacerbate different aspects of hydrogen embrittlement, depending on the balance between residual stress stabilization and microstructural susceptibility. This trade-off highlights the need to evaluate LSP-treated materials not only in terms of residual stress magnitude and penetration depth, but also with respect to hydrogen–microstructure interactions under realistic service loading, including cyclic and impact conditions.

In the literature, Laser Shock Peening (LSP), Ultrasonic Shot Peening (USP), and conventional shot peening are most commonly compared as near-surface modification techniques aimed at improving fatigue performance and resistance to environmental degradation. However, these methods differ fundamentally in both their interaction mechanisms and the scope of material changes they impose. LSP is typically associated with deeper penetration of compressive residual stresses and a lower risk of surface damage, whereas shot peening and USP often lead to a stronger increase in roughness and potential surface contamination due to direct contact with the impact media. These differences are directly relevant for hydrogen applications, where surface condition and chemical cleanliness can significantly influence hydrogen entry processes and damage initiation.

The costs and industrial implementation pathways of these technologies also differ. LSP requires specialized laser infrastructure and precise process control, which may limit accessibility for high-volume applications, while shot peening and USP are more widely established and often easier to implement industrially. In practice, the selection of a surface treatment therefore represents a compromise among targeted microstructural and stress-state effects, surface-quality constraints, and process economics.

Crucially, the main challenge in interpreting comparative data on hydrogen embrittlement performance is not necessarily a lack of results, but rather methodological limitations in many comparison studies. In most comparative investigations, the effects of surface roughness, surface contamination/foreign inclusions, and residual stress depth/profile are not decoupled in a rigorous manner. Consequently, observed differences in hydrogen embrittlement resistance may reflect side effects associated with surface condition or specimen preparation rather than intrinsic differences in the mechanism of a given treatment technology. This motivates comparative studies based on harmonized protocols in which the following are simultaneously controlled and characterized: (i) surface roughness and surface chemical state, (ii) residual stress profiles as a function of depth, and (iii) hydrogen exposure conditions together with the mechanical testing methods used to quantify degradation.

Overall, the available results and comparisons indicate that the effectiveness of impact-based treatments, including LSP, cannot be assessed using a single parameter such as peak compressive residual stress or stress penetration depth alone. Interactions among stress state, microstructure, and hydrogen transport and trapping are strongly coupled and depend on both the material system and the mechanical and environmental loading conditions. The differences reported between LSP, USP, and conventional shot peening further underscore that ambiguity in the literature often arises from methodological constraints, particularly the lack of systematic studies that concurrently control and quantify surface condition, residual stress profiles, microstructural evolution, and realistic hydrogen exposure. Without such an approach, it is difficult to separate beneficial effects associated with compressive stresses from potentially detrimental effects associated with trapping-driven hydrogen availability and localized plasticity.

Consequently, further progress in designing surface treatments resistant to hydrogen effects requires an integrated research approach that combines advanced characterization of residual stresses, microstructure, and hydrogen behaviour with mechanical-property evaluation under service-relevant conditions. This need leads directly to the identification of key application-oriented knowledge gaps, which are discussed in the subsequent part of this review. A separate and particularly promising direction is the use of LSP as a finishing treatment for parts refurbished by laser cladding. While laser cladding is central to circular-economy approaches (repair/remanufacturing), it can introduce critical defects such as tensile residual stresses and gas porosity. A landmark study by Wang et al. [40], published in *International Journal of Fatigue*, demonstrated that LSP can effectively reverse these detrimental effects in IN718. The authors identified a so called *defective pore closure effect*, whereby the shock wave mechanically closes near-surface voids while simultaneously transforming tensile residual stresses into high-amplitude compressive stresses. From the perspective of hydrogen embrittlement mitigation, this result is highly significant: porosity elimination removes volumetric hydrogen traps that without LSP can become primary sites for blistering and cracking. Wang showed that this synergy can translate into a pronounced extension of fatigue life, supporting the argument that LSP may become a critical step in remanufacturing workflows for hydrogen-service components.

In a critical comparison of mechanical surface treatments, Maurya and Akhtar [23] highlight a fundamental limitation of conventional shot peening in hydrogen applications: surface contamination. Residues of the peening media (e.g., embedded iron or ceramic particles) can form local galvanic couples which, in humid hydrogen environments, become centres for pitting corrosion and preferential hydrogen adsorption (promoted by cathodic reactions). LSP, being a non-contact process (typically employing an ablative sacrificial layer or operated in LPwC-type regimes), eliminates this risk. The authors emphasize that maintaining surface chemical cleanliness while introducing compressive stresses makes LSP a method of choice for sectors with strict cleanliness constraints (e.g., fuel cell components or composite tanks with metallic liners), where foreign inclusions are unacceptable.

Within the spectrum of modern surface strengthening technologies, Ultrasonic Shot Peening (USP) and Water Jet Peening (WJP) represent important alternatives to LSP. Comparative analysis in the context of hydrogen embrittlement mitigation, however, reveals fundamental differences in the depth and character of the induced modifications. While USP can yield smoother surfaces than classical pneumatic shot peening, its depth of influence (compressive stress depth) is typically limited to ~100–200 µm. By contrast, LSP routinely produces compressive layers on the order of ~1–2 mm, which can be a decisive advantage for protection against deeper hydrogen diffusion and subsurface cracking. Water Jet Peening (WJP), often applied in the nuclear sector to protect welds against stress corrosion cracking, similarly eliminates the risk of foreign inclusions. Nevertheless, the kinetic energy of water droplets is substantially lower than the plasma pressure generated in LSP (on the order of GPa), which makes WJP less effective for high-hardness, high-yield-strength materials such as modern maraging steels or nickel-based superalloys relevant to hydrogen infrastructure. In this sense, LSP remains a unique technology capable of introducing deep compression in ultra-high-strength materials without degrading surface stereometry, as emphasized by Maurya and Akhtar.

Taken together, these comparative observations indicate that the hydrogen-related performance of impact-based surface treatments cannot be inferred from a single descriptor (e.g., peak CRS or roughness) because stress profiles, surface condition, and trap populations are intrinsically coupled. This coupling motivates numerical support, particularly finite element analysis (FEA), as a means to design LSP process windows that maximize compressive protection while minimizing hidden tensile *weak zones* under hydrogen-relevant service conditions.

### 3.3. Numerical Support: Stress-Field Modelling and Risk Assessment (FEA)

The development of Laser Shock Peening (LSP) is increasingly intertwined with progress in numerical methods, as highlighted by the comprehensive review by Wakchaure, Misra, and Menezes [41]. The authors argue that finite element analysis (FEA) has evolved from a validation tool toward a key predictive instrument that is now essential for optimizing complex process parameters such as overlap ratio and pulse sequencing. Wakchaure et al. emphasize [42] that simulation reliability critically depends on the choice of an appropriate material constitutive model. In commercial environments such as ABAQUS or ANSYS, the Johnson–Cook (J–C) model is most frequently adopted because it can capture the extremely high strain-rate regime (often cited as $> 10^6 s^{-1}$) associated with laser-generated shock waves. The review also underlines the growing relevance of multi-shot peening simulations: interactions among successive stress waves are difficult to resolve experimentally, whereas numerical modelling enables systematic design of compressive residual stress (CRS) field uniformity.

A practical implementation of these principles is provided by the earlier-discussed work of Zhang et al. [33] on the MAR-M247 superalloy. Using an explicit Johnson–Cook (J-C) formulation, they not only reproduced the CRS profile but, crucially for hydrogen embrittlement (HE) mitigation, identified and minimized subsurface tensile residual stress (TRS) regions through optimization of laser spot size. Taken together, the review-level perspective (Wakchaure et al. [41]) and the application-driven demonstration (Zhang et al. [33]) point to an emerging trend toward digital twins of LSP processes. Such digital twins would enable virtual assessment of hydrogen-relevant performance at the component level prior to physical manufacturing, thereby reducing development cost and deployment risk. This shift is particularly important because trial and error optimization of LSP parameters is expensive and time-consuming, while high-fidelity numerical frameworks can explore process windows systematically.

Zhang et al. [33] presented a comprehensive finite element model based on the J-C constitutive equation, implemented in ABAQUS using the VDLOAD subroutine. Their simulations revealed a critical dualism for HE mitigation. On the benefit side, increasing laser power density from 9.63 to 14.45 $\mathrm{GW/cm^2}$ produced a near-linear rise in maximum surface compressive residual stresses (from 427 to 501 MPa) and increased the depth of compressive stress influence. In parallel, however, the numerical model precisely localized a potentially hazardous increase in subsurface tensile residual stresses within the force-balance region. Zhang showed that at higher laser power, the tensile stress peak not only becomes higher but also shifts deeper into the material. From a hydrogen perspective, this is a fundamental insight: TRS regions can act as *magnets* for hydrogen because tensile hydrostatic stress increases the local chemical potential driving hydrogen accumulation. Zhang's simulations therefore demonstrate that indiscriminately increasing laser power, while beneficial for surface hardening may create hidden subsurface regions prone to hydrogen-assisted crack initiation. The authors suggest that the solution lies in manipulating spot size: reducing the spot diameter while maintaining power density can limit both the depth and amplitude of detrimental tensile stresses. This provides a concrete design guideline for engineering a hydrogen-resistant near-surface layer, where the compressive barrier is maximized without introducing subsurface tensile *weak zones*. Summary of these investigations is presented in Table 1.

| **Aspect (FEA output)** | **Benefit when power density increases** | **Risk when power density increases** | **Hydrogen-relevant implication** |
|---|---|---|---|
| Surface CRS magnitude | Increases approximately linearly (427 → 501 MPa as power density rises from 9.63 to 14.45 $GW/cm^2$) | — | Stronger compressive barrier against hydrogen-assisted cracking initiation at/near surface |
| Depth of CRS influence | Increases (deeper compressive zone) | — | Delayed hydrogen penetration and improved resistance to subsurface damage, if tensile zones remain controlled |
| Subsurface TRS (force-balance region) | — | TRS peak increases and shifts deeper into the material | TRS regions can act as hydrogen "attractors" under tensile hydrostatic stress, creating hidden sites for hydrogen-assisted crack nucleation |
| Process-design lever | Spot-size optimization can maintain CRS benefits | Without spot-size control, higher power can amplify subsurface TRS | Reducing spot diameter (at constant power density) can limit TRS depth/amplitude while retaining compressive strengthening, providing a practical hydrogen-safe process window |

*Table 1 FEA derived benefit / risk trade-off for LSP parameter escalation in MAR-M247 (Zhang et al. [33]).*

In parallel to experimental progress, a notable advance is emerging in numerical modelling of trapping kinetics. A new full-field model introduced by Hussein et al. [42] enables simulation of hydrogen diffusion in two-phase microstructures while explicitly accounting for the spatial distribution of traps. Although the model was originally developed for duplex steels, its implications for LSP are direct: it allows prediction of thermal desorption spectroscopy (TDS) curve shapes as a function of charging time and defect density. Integrating such kinetics-based models with post-LSP stress maps obtained from FEM constitutes a promising direction for predicting hydrogen-barrier durability. In practice, this would enable virtual assessment of whether a given compressive layer will become saturated with hydrogen over a specified service time, thereby linking process parameters, microstructural defect populations, and long-term hydrogen exposure in a unified predictive framework.

### 3.4. Application-Oriented Knowledge Gaps

Despite the widespread use of slow strain rate tensile testing (SSRT) and fatigue crack growth studies in hydrogen embrittlement research, the literature still exhibits a significant deficit of engineering level knowledge regarding the effect of Laser Shock Peening (LSP) on critical safety parameters of materials operating in hydrogen environments. In particular, the available evidence remains insufficient when it comes to assessing the behaviour of LSP-treated materials under dynamic loading conditions, which are central to the structural integrity of hydrogen infrastructure components.

One of the most underexplored aspects is the combined effect of LSP and hydrogen exposure on impact toughness, typically evaluated using Charpy impact testing. Impact toughness quantifies a material's ability to absorb energy under sudden, high-rate loading and constitutes a key safety parameter for pressure vessels, pipelines, and related components; especially at reduced service temperatures that can reach approximately $-80°C$. Available studies indicate that pre-charging with hydrogen may cause a dramatic drop in absorbed impact energy, reaching up to ~50% at 77 K in austenitic steels such as 304L, whereas the effect at room temperature can be far less pronounced. From an operational safety perspective, these results represent an important warning signal.

In the context of Laser Shock Peening, the problem becomes even more complex. LSP introduces into the near surface region both deep compressive residual stresses, which may form an effective barrier to hydrogen transport, and an intense microstructural modification involving grain refinement and a substantial increase in dislocation and subgrain boundary density. In stable austenitic steels such as AISI 316L, grain refinement and increased dislocation density represent major strengthening mechanisms; however, they may simultaneously increase susceptibility to hydrogen embrittlement because high dislocation density correlates with enhanced hydrogen trapping and with changes in deformation mechanisms. This leads to a fundamental engineering conflict. On the one hand, LSP induced compressive residual stresses can effectively limit hydrogen diffusion and delay crack initiation. On the other hand, surface nanostructuring and the associated increase in hardness, often reported to reach ~30% after hydrogen exposure, may promote unfavourable deformation modes, such as planar slip, thereby reducing the ability of the material to absorb impact energy. This trade-off between strengthening and resistance to dynamic loading remains largely unresolved in the literature.

An additional interpretive limitation is that most studies of hydrogen embrittlement following LSP rely on electrochemical hydrogen charging. This approach often produces steep hydrogen concentration gradients and does not always faithfully represent real service conditions. Differences between internal and external hydrogen exposure further complicate direct comparison across experimental results.

Consequently, there is a clear need for systematic studies aimed at quantifying the impact of LSP on impact toughness and on material response under dynamic loading in hydrogen environments, particularly for high-strength steels intended for cryogenic and high-pressure service. Testing the hypothesis of a potentially adverse effect of LSP-induced nanostructuring on impact resistance in the presence of hydrogen is critical for ensuring safe and reliable deployment of hydrogen infrastructure.

Beyond this general picture, the literature reveals two principal application-oriented knowledge gaps concerning the behaviour of LSP-treated materials in hydrogen environments (summarized in Fig. 1). Both have direct implications for structural safety and have not been systematically addressed.

*Gap 1: Lack of quantitative structure–performance correlation under hydrogen (SSRT/fatigue).*

The first gap concerns the lack of correlation between mechanical test outcomes in hydrogen environments and detailed structural characterization of LSP-treated materials. Most available studies focus on hydrogen transport kinetics (e.g., permeation tests) and on basic tensile testing, including SSRT. While these methods provide valuable insights into embrittlement susceptibility, they do not fully represent realistic service conditions, where cyclic and fatigue loading dominates. In particular, SSRT despite its popularity has limited ability to simulate long-term operating conditions. It does not capture cumulative fatigue damage, nor does it reflect the time-dependent evolution of microstructure and residual stress fields that can be critical during sustained hydrogen exposure. As a result, the literature contains very few studies that quantitatively correlate residual stress state, dislocation density, and the degree of LSP-induced nanostructuring with fatigue crack growth parameters, such as $\Delta K$–$da/dN$ curves measured directly in hydrogen-containing environments. The absence of such studies prevents a clear assessment of whether LSP's protective effects persist under long-term cyclic loading.

*Gap 2: Neglect of impact toughness and long-term degradation mechanisms.*

The second gap is the near-complete omission of impact toughness in studies of hydrogen embrittlement for LSP-treated materials. Impact toughness is a core safety parameter, particularly in the context of sudden loads, vessel failures, or operation at reduced temperatures. Nevertheless, the literature provides virtually no systematic data on the influence of LSP on impact energy in the presence of hydrogen. Mechanistically, this is a particularly complex issue. The nanostructured zone produced by LSP may serve as an effective barrier to hydrogen transport, limiting diffusion into the bulk. Yet intense nanostructuring and increased defect density can also reduce the overall ability to absorb impact energy if the structure is not properly optimized. Hydrogen may amplify this effect by enhancing localized plasticity or promoting microcrack initiation. The conflict between increased strength and retained ductility/toughness remains unresolved and requires targeted experimental investigation.

The identified gaps clearly indicate that further development of LSP for hydrogen applications requires moving beyond laboratory tests of limited representativeness. Integrated studies are needed that combine quantitative characterization of microstructure and residual stresses with evaluation of fatigue and impact behaviour in the presence of hydrogen, enabling a credible assessment of safety and durability for hydrogen infrastructure components.

A key application challenge remains the optimization of pulse energy. Although increasing energy (e.g., from 1.2 J to 2 J) can generate deeper compressive residual stress fields, the accompanying strong rise in surface roughness (with $R_a$ reported to increase up to fourfold) becomes a limiting factor. Excessive roughness can act as a local stress micro-concentrator, accelerating hydrogen uptake and crack initiation, underscoring the need for precise parameter selection for specific infrastructure application [16].

Another important gap is the lack of standardized methods to assess hydrogen-induced degradation in small material volumes, which is essential for evaluating thin, surface-modified layers (e.g., those produced by LSP). A potential solution is the methodology proposed by Álvare et al. [5], based on the Notched Small Punch Test (SPT). They demonstrated that SPT can quantify reductions in fracture energy and ductility in hydrogen environments using miniature specimens. This technique enables selective assessment of gradient regions, such as the near-surface layer after laser peening, providing an advantage over classical macroscopic tests that effectively average the response across the entire specimen volume.

A critical methodological analysis by Asadipoor et al. [43] on X70 pipeline steel demonstrated fundamental differences in material response depending on the hydrogen charging mode. Their work showed that ex-situ testing (after pre-charging) can underestimate HE susceptibility because hydrogen can diffuse out of BCC specimens during the mechanical test, affecting primarily the surface layers. Only in-situ charging (continuous hydrogen supply during deformation) revealed the true quasi-cleavage fracture mechanism throughout the specimen volume. These results clearly indicate that credible evaluation of LSP barrier performance requires experiments conducted under continuous hydrogen exposure that better simulate real operating conditions.

A further gap in LSP implementation procedures for hydrogen infrastructure is the lack of integrated quality-control protocols that link residual stress assessment with topography analysis. Attolico et al. [30] emphasized that monitoring residual stress magnitude alone (e.g., by XRD or hole-drilling) is insufficient for assessing fatigue durability. In hydrogen environments, where surface adsorption processes play a key role, process-induced roughness must be treated as a critical parameter requiring optimization with the same rigor as the residual stress profile. Finally, Yu et al [18] noted that controversies remain regarding the effectiveness of surface treatments. They emphasized that although LSP introduces beneficial compressive stresses, it also generates defects (dislocations, grain boundaries) that can act as *diffusion highways* or crack initiation sites if not properly engineered. This underscores the need for precise process-parameter selection such that the benefits of the stress barrier outweigh the risks associated with increased defect density, an issue the authors describe in terms of optimizing *trap engineering*.

Taken together, these gaps indicate that advancing LSP toward hydrogen-service qualification requires a shift from predominantly coupon-scale, laboratory charging studies toward integrated, service-representative protocols that couple residual-stress and microstructure characterization with fatigue and impact performance under controlled hydrogen exposure thereby defining the priorities for future research and standardization discussed in the next section.

**Gap 1: Missing structure–performance correlation under hydrogen (SSRT/fatigue)**
Most studies report hydrogen transport/permeation and basic tensile metrics (including SSRT), but rarely quantify how *post-LSP residual stress fields and defect populations* translate into *hydrogen-relevant cyclic performance*.

- Lack of datasets linking CRS depth/profile + dislocation density/nanostructuring directly to $\Delta K – da/dN$ fatigue crack growth curves measured in $H_2$.
- SSRT alone is insufficient to represent service-relevant cyclic damage accumulation and time-dependent evolution of stress/defect fields.
- As a result, persistence of *LSP benefits* under long-term hydrogen-assisted fatigue remains uncertain.

**Gap 2: Impact toughness under hydrogen (Charpy) is largely unaddressed**
Systematic evidence for the combined effect of LSP + hydrogen exposure on impact energy is scarce, despite relevance to safety-critical components (pressure vessels, pipelines) and low-temperature service.

- Deep CRS may reduce hydrogen uptake/transport, while nanostructuring may either immobilize hydrogen or promote localized plasticity/microcrack initiation.
- The resulting strength–toughness trade-off in hydrogen (especially at cryogenic temperatures) is not resolved by the current literature.

**Cross-cutting methodological confounders (affecting both gaps):**

- Hydrogen charging mode: ex-situ vs in-situ; internal vs external hydrogen.
- Surface state: roughness/chemistry/contamination not decoupled from stress-depth effects.
- Limited suitability of conventional tests for thin gradient layers → need for miniature approaches (e.g., notched SPT) and harmonized protocols.

*Fig. 1 Application-oriented knowledge gaps in LSP-based mitigation of hydrogen embrittlement*

## 4. Conclusions and future research directions

Building on the coupled framework of hydrogen transport, trapping, and stress-field stability developed throughout this review, the following priorities outline the most urgent and the most enabling directions for advancing LSP toward reliable hydrogen-service qualification. The available evidence clearly indicates that a credible assessment of Laser Shock Peening (LSP) as a mitigation strategy for hydrogen embrittlement requires an approach that goes beyond conventional laboratory tests and simplified hydrogen-material interaction models. In particular, there is a pressing need for advanced in-situ characterization methods capable of tracking real-time changes both during LSP processing and during subsequent hydrogen exposure. The integration of synchrotron-based X-ray diffraction techniques such as EDDI and high-resolution HEXRD offers the opportunity to directly monitor the evolution and stability of residual stress fields under hydrogen charging. Such measurements would enable quantitative evaluation of stress relaxation and partial recovery dynamics and, critically, would clarify the extent to which LSP-induced compressive residual stresses remain effective in limiting hydrogen transport during service.

A major limitation of the current state of knowledge is the scarcity of long-term performance data for LSP-treated components operating in hydrogen-rich environments. Most studies focus on short-duration laboratory experiments, whereas engineering deployment requires evaluation under sustained cyclic loading (fatigue) and dynamic loading (impact/Charpy). Extending experimental campaigns to include fatigue in hydrogen and impact-toughness testing is essential for a reliable safety assessment of hydrogen infrastructure. Equally important is the lack of standardization across experimental methodologies used to evaluate hydrogen embrittlement after LSP. Variation in protocols, especially in SSRT and fatigue testing, severely limits cross-study comparability. Future research should therefore prioritize the development of harmonized, reproducible testing procedures representative of real service conditions, including fatigue testing in high-pressure gaseous hydrogen.

A further interpretive challenge stems from the widespread reliance on electrochemical hydrogen charging, which often produces steep hydrogen concentration gradients and may not represent real operating conditions. Differences between internal and external hydrogen exposure further complicate comparisons. As emphasized by Barthélémy, parameters such as gas pressure, exposure time, temperature, and specimen geometry can drastically influence outcomes and must be treated as first-order variables in experimental design. Overall, advancing LSP for hydrogen service demands an integrated framework combining stress and microstructure characterization with service-relevant mechanical testing under controlled hydrogen environments.

A particularly promising direction for resolving local hydrogen–microstructure interactions is the expanded use of micro-mechanical testing. Asadipoor et al. [43] successfully employed in-situ electrochemical micro-cantilever bending to demonstrate that hydrogen directly lowers yield stress at the microscale. Such approaches allow a more precise separation of hydrogen effects on plasticity (HELP-consistent behaviour) from decohesion-driven contributions, and they are especially well suited to thin, surface-modified layers produced by LSP, where macroscopic specimens cannot resolve steep gradients in properties.

While most current mitigation strategies focus on conventional steels, recent work also points to hydrogen as an active factor capable of modifying phase stability in advanced alloys. Wu et al. [44], studying complex concentrated refractory alloys (CCA), reported hydrogen-assisted spinodal decomposition, showing that hydrogen can dynamically alter thermodynamics and drive nanometre-scale chemical modulation and elemental segregation (e.g., Zr) under conditions where the alloy would otherwise be expected to remain stable. This has fundamental implications for surface engineering, including LSP, for hydrogen infrastructure: long-term hydrogen exposure may induce unexpected phase transformations within the near-surface layer that are not captured by classical diffusion-only models. Future LSP studies should therefore extend beyond stress analysis and incorporate how nanostructuring influences the thermodynamic stability of alloys under hydrogen.

A critical next step in validating LSP-based hydrogen barriers is the adoption of atomically resolved techniques. Chen et al. [20] note that conventional bulk methods (e.g., thermal desorption analysis) inherently average over the specimen volume and lose the spatial information required to confirm barrier mechanisms. A major methodological breakthrough is cryogenic atom probe tomography (Cryo-APT), which enables direct visualization of hydrogen (or deuterium) trapped at dislocations and grain boundaries. Applying Cryo-APT to LSP-modified surface layers could provide definitive evidence as to whether compressive stresses and defect architectures truly block hydrogen migration at the nanometre scale, one of the central unresolved questions in this field.

Closely related is the need for direct spatial mapping of hydrogen in modified surface layers. Traditional techniques such as TDS provide bulk hydrogen content but cannot resolve nanoscale distribution. Electron-beam methods (TEM/SEM) and many X-ray techniques (e.g., XPS) are generally ineffective for direct hydrogen detection because of hydrogen's low electron density and high mobility. In this context, ToF-SIMS has emerged as a powerful route for validating LSP barrier performance. Paudel et al. [45] highlighted ToF-SIMS as a technique capable of producing three-dimensional hydrogen concentration maps with nanometre-scale depth resolution. This enables direct assessment of whether a stress-based barrier truly arrests the hydrogen diffusion front or whether nanostructured interfaces behave as fast transport pathways.

Reliable hydrogen mapping, however, requires cryogenic protocols. Given hydrogen's high mobility in metals, room-temperature analysis may yield artefacts due to beam-induced migration. Performing measurements at approximately $-100°C$ or lower can *freeze* hydrogen within structural traps formed by LSP. Moreover, isotopic substitution using deuterium (D) reduces background from residual hydrogen in vacuum systems and enables precise tracking of ingress kinetics through the strengthened layer. Coupling ToF-SIMS with stress modelling therefore represents a missing link toward full validation of LSP technology for hydrogen applications.

Neutron-based methods also offer major advantages for thick-section components. The limited penetration depth of laboratory XRD (micrometre-scale) becomes a barrier to understanding LSP–hydrogen interactions in thick-walled structures. Cho et al. [22] highlight the strengths of in-situ neutron diffraction, which can map lattice strains and dislocation density in bulk volumes. Applying neutron diffraction to LSP-treated layers would help determine whether compressive stress relaxation under hydrogen is purely a surface phenomenon or extends deeper into the shock-affected zone. The integration of DLPA (diffraction line profile analysis) with hydrogen-mechanical testing is therefore a particularly promising pathway for validating barrier durability.

Beyond experimental methods, Yu et al. [18] suggest that the future of HE mitigation lies in coupling experiments with machine learning (ML) and digital twin concepts. ML-driven analysis of TDS datasets and microstructure mapping could yield predictive models for hydrogen-assisted fatigue durability after LSP, accounting for the complex coupling between residual stress profiles, trap populations, and local hydrogen concentration, potentially reducing reliance on expensive in-situ testing for every parameter permutation.

Important implementation questions remain, including the thermal stability of LSP-induced states. Cao et al. [39] highlight that elevated service temperatures can promote dislocation annihilation and relaxation of beneficial compressive stresses. For hydrogen infrastructure, this implies that LSP barrier effectiveness may degrade over time at higher temperatures. Process design should therefore consider stabilizing dislocation structures (e.g., via precipitate pinning) to prevent premature microstructural recovery and loss of hydrogen-trapping capability.

A further emerging trend is the hybridization of surface-engineering strategies. Maurya and Akhtar [23] argue that mechanical treatments alone (such as LSP) may be insufficient in extreme hydrogen environments and propose sequential systems in which LSP is used as a pre-treatment prior to applying barrier coatings (e.g., PVD or DLC). In such architectures, LSP stabilizes the substrate by introducing compressive stresses and increased hardness, reducing substrate plasticity beneath the hard coating and mitigating coating cracking or hydrogen blistering. This approach may be particularly promising for next-generation hydrogen components requiring both chemical barrier performance and mechanical robustness.

Finally, a recent synthesis by Huang et al. [24] provides valuable insight into fatigue behaviour after LSP in hydrogen environments, an area still weakly resolved. The authors propose that LSP can act in two complementary ways: delaying fatigue crack initiation through deep compression (closing near-surface microcracks), while also modifying crack-growth kinetics. In hydrogen, they argue that a nanostructured near-surface layer (on the order of ~20 – 50 µm) can function as a *delay shield*, reducing crack growth rate through increased tortuosity and trapping at grain boundaries. However, the benefit is limited by the depth of the compressive layer; once the crack propagates beyond this region, acceleration can occur. This supports the need for deep LSP strategies (e.g., multiple impacts) in safety-critical components where hydrogen-assisted crack growth is the governing failure mode.

The conclusions from above are presented and summarized in Fig. 2.

Near-term priorities (qualification-oriented)

- Service-representative testing: fatigue in high-pressure gaseous hydrogen and impact toughness (Charpy) under hydrogen, including low-temperature regimes where relevant.
- Protocol harmonization: standardized SSRT/fatigue/charging procedures with clear reporting of pressure, temperature, exposure time, and specimen geometry (internal vs external hydrogen).
- Integrated quality control: coupled assessment of residual-stress profiles + surface topography/chemistry (roughness and contamination treated as critical variables).
- Miniature methods for gradient layers: notched SPT and micro-mechanical testing to isolate near-surface LSP layers and resolve property gradients.

Enabling technologies (mechanism-resolving & predictive):

- In-situ diffraction under hydrogen: synchrotron EDDI/HEXRD and in-situ neutron diffraction + DLPA to quantify residual-stress evolution and defect dynamics in bulk.
- Nanoscale hydrogen mapping: cryogenic protocols + isotope tracing (D); 3D profiling with ToF-SIMS to verify whether LSP truly arrests hydrogen fronts.
- Atomic-scale validation: Cryo-APT to directly localize H/D at dislocations, grain boundaries, and precipitates within LSP-modified layers.
- Digital twins & ML integration: combined FEM stress maps + trapping/transport models + data-driven inference to predict barrier durability and hydrogen-assisted fatigue life.

*Fig. 2 Priorities for advancing LSP-based mitigation of hydrogen embrittlement: near-term actions vs enabling technologies*

In summary, translating LSP from a promising laboratory-scale intervention into a hydrogen-service qualified technology requires treating residual stresses, microstructural trapping, and surface state as a coupled system that evolves during hydrogen exposure and under realistic loading. The most urgent need is not additional isolated datasets, but harmonized, service-representative protocols that combine stress/microstructure characterization with fatigue and impact performance in controlled hydrogen environments. In parallel, emerging in-situ, cryogenic, and atomically resolved techniques (e.g., EDDI/HEXRD, neutron diffraction with DLPA, ToF-SIMS, and Cryo-APT) provide the missing capability to directly validate where hydrogen resides and how barriers fail. Finally, integrating these experimental insights with physics-informed modelling and digital-twin frameworks offers a practical route to optimizing LSP process windows while reducing development cost and risk for hydrogen infrastructure deployment.

**Acknowledgements**

Co-funded by the European Union (MERIT - Grant Agreement No. 101081195). Views and opinions expressed are, however, those of the authors only and do not necessarily reflect those of the European Union or the Central Bohemian Region. Neither the European Union nor the Central Bohemian Region can be held responsible for them.

This work was co-funded by European Union and the state budget of the Czech Republic under the project LasApp CZ.02.01.01/00/22_008/0004573.

**CRediT authorship contribution statement**

Elżbieta Gadalińska: Conceptualization; Literature search; Data curation; Writing – original draft; Writing – review & editing; Visualization; Supervision; Project administration

Jan Kaufman: Writing – review & editing; Validation.

Jan Šmaus: Writing – review & editing; Validation.

Jan Brajer: Writing – review & editing; Validation.

**Declaration of generative AI and AI-assisted technologies in the manuscript preparation process**

During the preparation of this work, the authors used ChatGPT (OpenAI) for language editing, structural refinement, and improving the clarity of the manuscript. After using this tool, the authors reviewed and edited the content as needed and take full responsibility for the content of the published article.